\documentclass[prl,twocolumn,showpacs,superscriptaddress,floatfix]{revtex4}
\usepackage{graphicx}
\newcommand{\ket}[1]{\left| #1 \right\rangle}
\newcommand{\bra}[1]{\left\langle #1\right |}
\begin{document}
\title{Aperiodic Quantum Random Walks}

\author{P. Ribeiro$^1$, P. Milman$^2$ and R. Mosseri}

\address{$^1$Groupe de Physique des Solides, Universit\'{e}s Paris 6 et 7,\\
 campus Boucicaut, 140 rue de Lourmel, 75015 Paris France. \\
 $^2$ Laboratoire de Recherche en Informatique, Université Paris-Sud, 91405 Orsay, France}
\email{pedro.ribeiro@polytechnique.fr}
\email{milman@lri.fr}
\email{mosseri@gps. jussieu.fr}

\date{\today}

\begin{abstract}
We generalize the quantum random walk (QRW) protocol for a particle in a one-dimensional chain, by using several types of biased quantum coins, arranged in aperiodic sequences, in a manner that leads to a rich variety of possible wave function evolutions. Quasiperiodic sequences, following the Fibonacci prescription, are of particular interest, leading to a sub-ballistic wavefunction spreading. In contrast, random sequences leads to diffusive spreading, similar to the classical random walk behaviour. We also describe how to experimentally implement these aperiodic sequences.   

\end{abstract}

\maketitle

A quantum random walk (QRW) is a natural extension to the quantum world of the
ubiquitous classical random walk. It was first proposed in \cite{LUIZ:PRA93}
and widely investigated recently (see the recent review by Kempe \cite{KEMPE:review}), mostly in connection with possible
applications to quantum algorithms \cite{KEMPE:PRA03,AMBIANIS}. The generic discrete QRW consists in a  particle moving on a graph , in a direction depending on its internal state (either called spin or chirality), the simplest case being a spin  ${\frac{1}{2}}$ particle on a periodic chain.
In between each moving step, an unitary transformation, called a ``quantum coin operator" (QCO), is acted on the particle spin state and shuffles the spin related amplitude of the particle wave function. Most studies have been focused on the so-called Hadamard transform, but more general QCO can been used. 
A main difference between the classical and quantum walks is seen on the particle spreading,
 as measured by the long time dependence of the standard deviation  $\sigma(t)=\sqrt{\left\langle
x^{2}\right\rangle_{t} -\left\langle x\right\rangle_{t} ^{2}}.$ The classical case
displays a diffusive behavior ( $\sigma(t)\sim t^\frac{1}{2}$) while the quantum case 
is ballistic $\left( \sigma(t)\sim t\right)$, as can be proved in one dimension from the exactly computed solution \cite{AMBIANIS_01}. The latter result relies on the space periodicity of the QRW process, which allows for a  Fourier transformed wave function simple form. This reminds the behavior of tight-binding Bloch electrons under standard quantum evolution on a periodic lattice. It is therefore tempting to check whether well-known effects of quasiperiodicity in the latter case (like a sub-ballistic scaling with time of the standard deviation\cite{Abe_87}, or the  auto-correlation function \cite{Zhong_95}) can also be observed in QRW. We address this question here by generalising the QRW to the case where different quantum coins are applied along three types of sequences, either periodic, quasi periodic and random. 

The particle displacement is along a one dimensional periodic chain indexed by $k\in \textbf{Z}$,
 with a corresponding orthonormal basis $\left\{  \left\vert k\right\rangle \right\}  $ spanning the position Hilbert space $H_{P},$. To the quantum coin part corresponds a two-dimensional Hilbert space $H_{C}$ spanned by $\left\{  \left\vert \uparrow\right\rangle ,\left\vert \downarrow\right\rangle \right\}  $. The
particle wave function reads
\begin{equation}
\left\vert \Psi\right\rangle ={\sum_{s,k}}a\left(  s,k\right)
\left\vert s\right\rangle \otimes\left\vert k\right\rangle \text{
\ \ \ \ \ \ with }s\in\left\{  \uparrow,\downarrow\right\}
\end{equation}
The QRW unitary step operator $S(\alpha)$ is the concatenation of a displacement $D$  which reads
\begin{equation}
D={\sum_k}\left(  \left\vert \uparrow\right\rangle \left\langle
\uparrow\right\vert \otimes\left\vert k+1\right\rangle \left\langle
k\right\vert +\left\vert \downarrow\right\rangle \left\langle \downarrow
\right\vert \otimes\left\vert k-1\right\rangle \left\langle k\right\vert
\right)
\end{equation}

following a quantum coin operator $C(\alpha)$, a unitary transformation acting on the spin sector. Here we mainly use the rather simple QCO,

\begin{equation}
C\left(  \alpha\right)  =\left(
\begin{array}
[c]{cc}
\cos(\alpha) & \sin(\alpha)\\
\sin(\alpha) & -\cos(\alpha)
\end{array}
\right)  \label{genhad}
\end{equation}
which depends on a single parameter $\alpha\in\left[  0,\frac{\pi}{2}\right]$. Note that $C(\alpha)$ squares to the identity transformation.

\begin{figure}[h]

			\includegraphics[width=7cm]{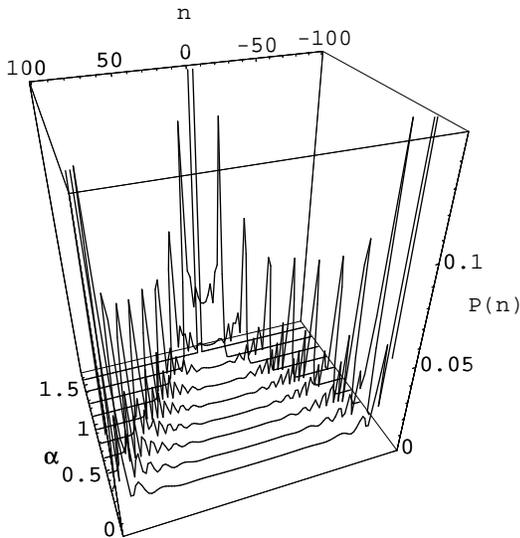}
\caption{Probability distributions after 100 time steps for 10 equally spaced $\alpha$ values between $0$ and $\pi/2$.  The middle curve corresponds to the standard Hadamard case. For $\alpha=0$, the QCO is simply $\sigma_z$, and the quantum walk amounts to two unrelated, left-right, ballistic moves, whose probability distribution is not distinguished from the bounding box vertical edges. For $\alpha=\pi/2$, the QCO is $\sigma_x$, and the quantum walk is confined near the origin}	
	\label{Gen_Had}
\end{figure}

For $\alpha=\pi/4$, one recovers the widely studied Hadamard walk, with its ballistic behavior. Other values of $\alpha$
also also gives rise to the same behavior, with different prefactors, except for $\alpha =\pi/2$ (corresponding to the $\sigma_{x}$ Pauli matrix) for which the particle remains confined. Note that for $\alpha=0$, the
QCO reduces to the diagonal Pauli matrix $\sigma_{z}$. In this case, the motion is 
completly decoupled into ballistic (right going) spin up and (left going) spin down parts,
 the latter acquiring a $\pi$ phase at each step. Figure \ref{Gen_Had} displays the typical wave function speading for several $\alpha$ values, with an initial ket $(\left\vert \uparrow\right\rangle+i\left\vert \downarrow\right\rangle)/\sqrt{2}$, which allows for a symmetric probability distribution. 

 Let us now combine two different step operators $S(\alpha)$ and $S(\beta)$ into infinite sequences and compute the
wave function spreading as measured by the exponent $c\left(  \alpha,\beta\right)$ in $\sigma\sim t^{c(\alpha,\beta)}$. For periodic sequences, the long time behaviour is still found to be ballistic (see below for quasiperiodic approximants), although displaying a more complex structure at the scale of one period. This suggests that new behaviours can be expected when the period size tends to infinity, as for the quasiperiodic case that we now study.

We consider  a Fibonacci sequence, obtained by iteration of the recursive rule $S_{n+1}=S_nS_{n-1}$, with $S_0=S(\alpha)$ and $S_1=S(\alpha)S(\beta)$. Given a sequence $S_n$, the next one can also be otained using a
substitution rule $S(\alpha)\rightarrow S(\alpha)S(\beta)$, $S(\beta)\rightarrow S(\alpha)$.
Only writing the $\alpha$ and $\beta$ symbols, the first sequences, called approximant sequences, read: $\alpha, \alpha\beta, \alpha\beta\alpha, \alpha\beta\alpha\alpha\beta, \alpha\beta\alpha\alpha\beta\alpha\beta\alpha \ldots$. The infinite sequence is not periodic. Indeed the occurence ratio of $\alpha$ versus $\beta$ tends to the  irrational golden mean $\tau=(1+\sqrt{5})/2$, which cannot be displayed in the repeated unit cell of a periodic sequence. This sequence is well known to be  quasiperiodic, a kind of order that has been widely investigated in the last 20 years in the context of quasicrystals. The effect of a sequence of quasiperiodic unitary tranformations  applied to a spin $1/2$ has been studied by Sutherland \cite{sutherland_86} and displayed a rich behaviour. The latter transformations were more generic than the simple QCO used here, but were not coupled to a displacement, as in the QRW. In the space sector, this sequence, when coding a quasiperioc potential, is well known to cause a sub-ballistic behaviour for tight-binding electrons \cite{Abe_87},\cite{Zhong_95}. In contrast, periodic repetition of the approximant sequences eventually leads to a ballistic spreading. 

\begin{figure}
	\begin{center}
		\includegraphics[width=8cm]{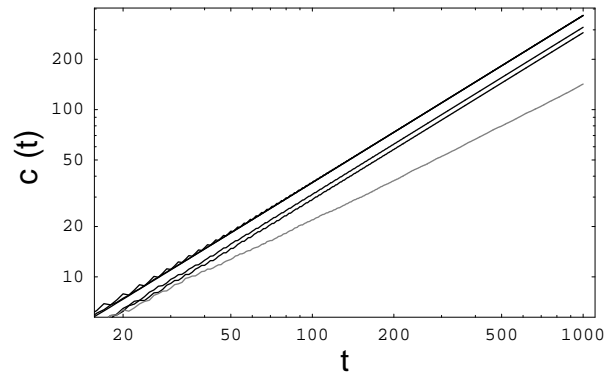}
	\end{center}
	\caption{Standard deviation $\sigma(t)$ of the probability distributions for periodically repeated approximant sequences (with period length of respectively 2,3,5 and 8), and for the asymptotic Fibonacci quasiperiodic sequence (grey colour). The latter clearly displays a sub-ballistic slope. These curves are obtained with the parameters ($\alpha=\pi/3,\beta=\pi/6$) }
	\label{spd}
\end{figure}

A similar behaviour is found here for the quasiperiodically shuffled QRW (with generic values of $\alpha$ and $\beta$). Recall that the position space is here periodic, and that the quasiperiodic modulation is applied with time
in the spin sector. The simulation was done for typically several thousands random walk steps. Let us first compare the spreading with time for periodically repeated small approximants and for the asymptotic quasiperiodic sequence. By the latter, we mean that we numerically generate a long sequence whose length is larger than both the chain length and the number of random walk steps. The standard deviation is plotted in figure \ref{spd}, which clearly displays a qualitative difference between the (asymptotically linear with time) periodic case and the slower quasiperiodic case.

\begin{figure}[h]
	
		\includegraphics[width=8cm]{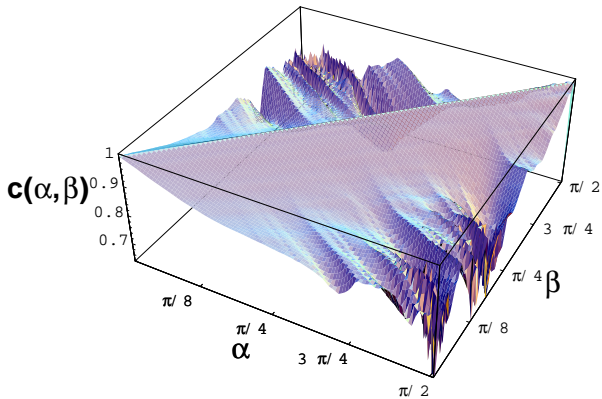}

	\caption{Slope of the standard deviation $\sigma(t)$, versus the parameters ($\alpha, \beta$) for a Fibonacci sequence.  }
	\label{fibonacci3D}
\end{figure}

The sub-ballistic behaviour is generic in the quasiperiodic case, whatever $\alpha$ and $\beta$ are. But the asymptotic slope $c(\alpha,\beta)$ is not a smooth function, as seen on figure \ref{fibonacci3D}. 
The ''diagonal " $\alpha=\beta$ corresponds to the periodic case and therefore to the expected ballistic slope $c=1$.
More surprising, and not yet fully understood, are the clearly visible transverse crest lines, whose  (equal) inverse slope is very close to the golden mean. To check whether this apparent arithmetical relation was not a simple coïncidence, we tried an alternative quasiperiodic sequence, based on the `silver' mean : $1+\sqrt{2}$. Parallel crest lines whose slope are simply related to the silver mean were again found. We are then tempted to appeal to the subtle properties of some iterated maps based on these quasiperiodic sequences \cite{kohmoto_83}, displaying either periodic (in the order of the Fibonacci sequence) or chaotic behaviour which are clearly seen in the computed quasiperiodically rotated spin system \cite{sutherland_86}. The above Fibonacci quantum coin sequence proves to have a simple cyclic behaviour for any value of the pair $(\alpha,\beta)$. But these crest lines also appear if we replace the aboce QCO (expression (\ref{genhad})) by a simple 2 dimensional rotation matrix, in which case the quasiperiodically rotated spin system is not generically cyclic.

Note that the QRW possible experimental implemantation (see below) is of high interest in the
context of quasiperiodic systems, since, in that case, the sub-ballistic behaviour, although clear from many 
computations, has not been yet clearly demonstrated in experiments.
 
\begin{figure}[h]
	
		\includegraphics[width=8cm]{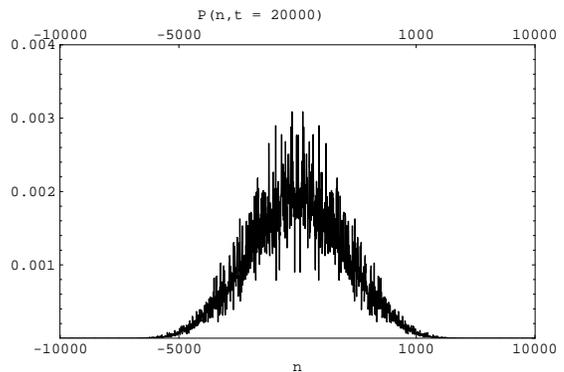}
	\caption{Typical probability distribution in the "`continuous"' random case. A "`hairy"'-lige gaussian shape is found, with a diffusive ($c\approx0.5$) standard deviation }	
	\label{gauss_1}
\end{figure}

Let us finally consider random sequences. We are still willing to compare with tight-binding electron evolution. In the latter case, a one-dimensional random potential is expected to generate a (very) long time wave packet localisation, preceeded, at short time, by a ballistic motion whose range depends on the error range width. We consider here two types of disordered QRW. We first generate $50-50$ random sequences with two different QCO, defined by $\alpha, \pi/2-\alpha $. To each random sequence corresponds a definite wave-function spreading, and we therefore average over many disorder realisations to check for an asymptotic regime. The second case under consideration is that of a continuous set of QCO, whose distribution, centred on $\alpha=\pi/4$, has variable width. 
This situation is related, but not identical, to a model of errors for an experimental implementation, which has been considered in \cite{DUR:PRA02} for periodic QRW in an optical lattice. 
In both cases, a diffuse regime, $c\approx 0.5$ is found, and the probability distribution have a gaussian-like shape with short range structure (figure \ref{gauss_1}).

\paragraph{Experimental implementation:}
We can now discuss how the above computed behaviours could be experimentally testet. There are, up to now, several experimental proposals for the realization of the usual QRW, among which, for example, one atom in an  optical lattice \cite{DUR:PRA02}, trapped ions \cite{TRAVAGLIONE:PRA02}, systems using linear optics  \cite{JEONG:PRA04} and cavity quantum electrodynamics \cite{LUIZ:PRA93,SANDERS:PRA03}. A continuous time version of the algorithm performed in a circular chain has already been realized using nuclear magnetic resonance \cite{DU:PRA03} and the difference with the classical random walk demonstrated. The realization of the aperiodic QRW involving biased alternating coins, represent only limited modifications to most of these protocols, implying that there is no major difficulty to the experimental implementation of the present aperiodic two coins sequences. We can study the specific example of optical lattices and describe how the protocol of aperiodic QRW can be implemented therein. In the proposal presented in \cite{DUR:PRA02}, the internal states of a Rb neutral atom (considered as a two state system) are subject to the quantum coin. This atom is located initially at one site of an one dimensional lattice, and will move to one of its neighbouring site according to its internal state. The conditional translation is performed in the same way as proposed in \cite{CIRAC}: different internal states feel different optical potentials and they are kept in the ground state  of their respective potential. For the case studied in \cite{DUR:PRA02}, the relevant internal atomic states are the hyperfine structure states $\ket{F=1,m_f=1}$  and $\ket{F=2,m_f=2}$. They will be denoted from now on as qubits $\ket{0}$ and $\ket{1}$.  The corresponding potential is $V_0(x,\theta)=[V_{m_s=1/2}(x,\theta)+3V_{m_s=-1/2}(x,\theta) ]/4$ and $V_1(x,\theta)=V_{m_s=1/2}(x,\theta)$, where $V_{m_s=\pm1/2}(x,\theta)=\alpha |E_0|^2\sin{(kx\pm\theta)}$ \cite{CIRAC}. The angle $\theta$ is half the angle between the polarizations of the two lasers that form the lattice and $E_0$ is the amplitude of the electrical field. For the specific choice of $\theta=0$, the minima for both hyperfine states coincide. The parameter $\theta$ can be adiabatically changed by turning one of the lasers polarization. The potential wells corresponding to each one of the internal states are thus translated with respect to each other. Since the minima for both internal states coincide again at $\theta=\pi$, at this point the two possible internal states are present in the same position.  The ``mixing" step, corresponding to the Hadamard gate, is performed by a laser tunned to the frequency separating the two atomic internal states. The application of the laser pulse occurs when the minima for both states coincide, and its duration determines the superposition of the two internal states that is created. The hamiltonian corresponding to the atom-laser interaction can be written as:
\begin{equation}
\hat H_{int}=\Omega (\ket{0}\bra{1}e^{i\phi}+\ket{1}\bra{0}e^{-i\phi},
\end{equation}
where $\Omega $ is the  Rabi frequency, $e^{i\phi}$ is the pulse's phase. The time evolution reads :
\begin{eqnarray}
\ket{0}(t)=\cos{(\Omega t)}\ket{0}-\sin{(\Omega t)}e^{-i\phi}\ket{1} \nonumber \\
\ket{1}(t)=\sin{(\Omega t)}e^{i\phi}\ket{0}+\cos{(\Omega t)}\ket{1}.
\end{eqnarray}
This means that an appropriate choice of the laser's phase and of the pulse duration can build all possible superpositions of the two states.

The modifications to the usual QRW protocol suggested in this paper demand changing only this point of the experimental proposal: the generalized coins  create biased superpositions of the internal states, and their realization is possible by controlling the laser pulse duration. In order to alternate two different GCO, one only needs to alternate two different pulse durations.

\paragraph{Conclusion:}

We have shown how the standard Quantum Random Walk framework can be considerably enriched by allowing more than one type of Quantum Coin Operator, arranged along different squences. In particular, quasiperiodic binary sequences lead to a sub-ballistic wave packet spreading characterised by sequence dependant slopes. The quantum state evolution depends on both the precise values taken by the two coins and the sequence itself. Note that we have studied here the simplest cases, with Fibonacci sequences and generalized Hadamard coins. More complex evolution are expected if the two coins are picked in a more generic set, or arranged along more complex (even not random) sequences. 
We have also studied random sequences, which leads to diffuse spreading, in contrast with the localisation effect encountered for quantum particles subjected to disordered potentials in one dimension. 

A very interesting aspect of these binary quantum random walk is their possible experimental implementation. Indeed sub-balistic spreadings are often computed (specially in the quasiperiodic case), but rarely observed. We have described here the mofication implied by going from single to binary sequences, which is not in principle very complicated.  

Acknowledgements: We would like to thank C. Aslangul and J. Vidal for several and fruitful discussions and comments



\begin{thebibliography}{9}                                                                                                %


\bibitem{LUIZ:PRA93} Y. Aharonov, L. Davidovich and N. Zagury, Phys. Rev. A {\bf 48}, 1687 (1993). 

\bibitem{KEMPE:review} J. Kempe, Contemporary Physics,{\bf 44}, 302327 (2003).

\bibitem {KEMPE:PRA03}N. Shenvi, J. Kempe and K. Brigitta Whaley, Phys. Rev.
A {\bf 67}, 052307 (2003).     

\bibitem{AMBIANIS} A. Ambianis, quant-ph/0305170 (2003); 



\bibitem{AMBIANIS_01} A. Ambianis, E. Bach, A. Nayak, A. Wishwanath and J. Watrous, \textit{Proceedings of the 33rd STOC}((Assoc for Comp. Machinery, New York, 2001);    
   
\bibitem{Abe_87} S. Abe and H. Hiramoto, Phys. Rev. A {\bf 36}, 5349 (1987)

\bibitem{Zhong_95} J.X. Zhong and R. Mosseri, J. Phys.:Condens. Matter {\bf 7}, 8383-8404 (1995). 

\bibitem{sutherland_86} B. Sutherland, Phys. Rev. Lett. {\bf 57}, 770 (1986)

\bibitem{kohmoto_83} M. Kohmoto, L.P. Kadanoff and C. Tang, Phys. Rev. Lett. {\bf 50}, 1870 (1983)
       
\bibitem{DUR:PRA02}
       W. D\"ur, R. Raussendorf, V. M. Kendon and H.-J. Briegel, Phys. Rev. A {\bf 66}, 052319 (2002).
\bibitem{SANDERS:PRA03} B. C. Sanders {\it et al.}, Phys. Rev. A {\bf 67}, 042305 (2003).

\bibitem{TRAVAGLIONE:PRA02} B. C. Travaglione and G. J. Milburn, Phys. Rev. A {\bf 65}, 032310 (2002).

\bibitem{JEONG:PRA04} H. Jeong, M. Paternostro and M. S. Kim, Phys. Rev. A {\bf 69}, 012310 (2004); M. Hillery, J. Bergou and E. Feldman, Phys. Rev. A {\bf 68}, 032314 (2003).




\bibitem{CIRAC} D. Jakcsh {\it et al.}, Phys. Rev. Lett. {\bf 82}, 1975 (1999); H.-J. Briegel {\it et al.}, J. Mod. Opt. {\bf 47}, 415 (2000).      



\bibitem{DU:PRA03} Jiangfeng Du, Hui Li, Xiaodong Xu, Mingjun Shi, Jihui Wu, Xianyi Zhou, and Rongdian Han , Phys. Rev. A 67, 042316 (2003)

\end{thebibliography}
\end{document}